\documentclass[submission,copyright,creativecommons]{eptcs}
\providecommand{\event}{SR 2013} 
\usepackage{latexsym,amssymb,theorem,fleqn}

\newcommand{\LTL}{\mathit{LTL}}
\newcommand{\CTL}{\mathit{CTL}}

\newcommand{\ATL}{\mathit{ATL}}
\newcommand{\ATLKP}{{\mathit{ATL}^{D}_{iR}}}

\newcommand{\fin}{\mathit{fin}}

\newcommand{\atlD}[1]{\langle\!\langle#1\rangle\!\rangle}
\newcommand{\atlB}[1]{[\![#1]\!]}

\newcommand{\yesterday}{\mathbf{\ominus}}

\newcommand{\until}[2]{{({#1}{\mathsf U}{#2})}}
\newcommand{\wuntil}[2]{{({#1}{\mathsf W}{#2})}}

\newcommand{\since}[2]{{({#1}{\mathsf S}{#2})}}

\newcommand{\dknows}{{\mathsf{D}}}
\newcommand{\poss}{{\mathsf{P}}}
\newcommand{\Act}{{\mathit{Act}}}

\newcommand{\out}{{\mathrm{out}}}

\newcommand{\oomit}[1]{}

\newtheorem{all}{Proposition}
\newtheorem{lemma}[all]{Lemma}
\newtheorem{proposition}[all]{Proposition}

{\theorembodyfont{\upshape}\newtheorem{definition}[all]{Definition}}
{\theorembodyfont{\upshape}}
\newenvironment{proof}{\noindent
{\bf Proof:}}{$\dashv$}
\newenvironment{proof*}{\noindent
{\bf Proof:}}{$\dashv$}

\title{Reducing Validity in Epistemic ATL\\ to Validity in Epistemic CTL}
\author{Dimitar P. Guelev
\institute{Institute of Mathematics and Informatics\\
Bulgarian Academy of Sciences\\
Sofia, Bulgaria}
\email{gelevdp@math.bas.bg}
}
\def\titlerunning{Reducing Validity in Epistemic ATL}
\def\authorrunning{D. P. Guelev}
\begin{document}
\maketitle

\begin{abstract}
We propose a validity preserving translation from a subset of epistemic Alternating-time Temporal Logic ($\ATL$) to epistemic Computation Tree Logic ($\CTL$). The considered subset of epistemic $\ATL$ is known to have the finite model property and decidable model-checking. This entails the decidability of validity but the implied algorithm is unfeasible. Reducing the validity problem to that in a corresponding system of $\CTL$ makes the techniques for automated deduction for that logic available for the handling of the apparently more complex system of $\ATL$. 
\end{abstract}

\section*{Introduction}

\oomit{
Systems of {\em Alternating time temporal logic} ($\ATL$, \cite{AHK97,AHK02}) vary on the restrictions on the considered games such as the players' information on the game state, which may be either {\em complete} or {\em incomplete} ({\em imperfect}), and their ability to keep complete record of the past, which is known as {\em perfect recall} \cite{JH04,Sch04,DBLP:conf/ijcai/JamrogaB11}. A classification of the variants of epistemic linear- and branching-time {\em temporal} logics (without the game-theoretic modalities) can be found in \cite{MeydenWong03,HMV2004}. 
}

The strategic cooperation modalities of {\em alternating time temporal logic} ($\ATL$, \cite{AHK97,AHK02}) generalize the path quantifier $\forall$ of {\em computation tree logic} ($\CTL$). Combinations of $\ATL$ with modal logics of knowledge \cite{HW03,JH04} extend temporal logics of knowledge (cf. e.g \cite{HFMV95}) in the way $\ATL$ extends $\CTL$. Automated deduction for $\CTL$ and linear time epistemic temporal logics has been studied extensively \cite{DBLP:journals/tocl/FisherDP01,DBLP:conf/mfcs/BolotovDF99,DBLP:conf/atal/GorankoS09,DBLP:conf/mallow/GorankoS09}. There is much less work on the topic for $\ATL$, and hardly any for its epistemic extensions. The decidability of validity in $\ATL$ with complete information was established in \cite{GD06} as a consequence of the {\em finite model property}, where the completeness of a Hilbert-style proof system was given too. Hilbert-style proof systems are known to be unsuitable for automating proof search. The situation was remedied by a tableau-based decision procedure developed in \cite{DBLP:journals/tocl/GorankoS09}. Along with that, the same authors developed tableau systems for branching epistemic temporal logics in \cite{DBLP:conf/mallow/GorankoS09}. Temporal resolution (cf. e.g. \cite{DBLP:journals/tocl/FisherDP01}), which is well understood for linear time logics and their epistemic extensions, was considered for $\ATL$ in \cite{Lan2010}, but only for the $\atlD{.}\circ$-subset, which is similar to {\em coalition logic} \cite{Pauly02} and enables only reasoning about a fixed number of steps. To our knowledge, no similar work has been done for systems epistemic $\ATL$.

In this paper we continue the study \cite{GuelevDE11} of a system of $\ATL$ with the operator of distributed knowledge under the perfect recall assumption. In \cite{GuelevDE11} we established the finite model property for a subset, and a model-checking algorithm for the whole system. That algorithm assumed that coalition members can use the distributed knowledge of their coalitions to guide their actions. Dropping that assumption is known to render model-checking undecidable \cite{DBLP:journals/corr/abs-1102-4225}. As expected, the validity-checking algorithm which these results imply is unfeasible. 

In this paper we propose a validity preserving translation from another subset of that logic into epistemic $\CTL$, with distributed knowledge and perfect recall again. As it becomes clear below, the need to consider a subset appears to be due to the lack of connectives in epistemic $\CTL$ to capture some interactions between knowledge and the progress of time. The translation makes no assumption on coordination within coalitions and there is no dependence on the availability of the past temporal modalities which are featured in the axiomatization from \cite{GuelevDE11}. A semantic assumption that we keep is {\em finite branching}: only finitely many states should be reachable in one step from any state and models should have only finitely many initial states. Dropping that assumption would disable the fixpoint characterization of $\until..$-objectives that we exploit, because of the requirement on strategies to be uniform. The translation enables the use of the known techniques for mechanized proof in the apparently simpler logic $\CTL$ and its epistemic extensions \cite{DBLP:journals/jetai/BolotovF99,DBLP:conf/mallow/GorankoS09}.  Building on our previous work \cite{GuelevDE11}, we work with the semantics of $\ATL$ on {\em interpreted systems} in their form adopted in \cite{LR06a}. 

\oomit{
\noindent{\bf Structure of the paper} We present our translation by giving the translation step for $\ATL$'s $\atlD{.}\until..$-formulas first. That produces some fresh occurrences of the combination of these $\ATL$ modalities with $\circ$. Then we show how to eliminate $\atlD{.}\circ$ too, finally producing an epistemic $\CTL$ formula. Finally we discuss the cause for restricting our approach to a subset in the case of imcomplete information and give the translation clause for the excluded part of the language for the case of {\em complete information} $\ATL$, where no such obstacle occurs. Finally, we  review the work on automating deduction in systems related to epistemic $\CTL$ with distributed knowledge and perfect recall, which is the target logic of our translation. Brief preliminaries on $\ATL$, its semantics on interpreted systems, the considered subset, and the target epistemic extension of $\CTL$ are given in an appendix. 
}

\section{Preliminaries}
\subsection{Propositional epistemic $\ATL$ with perfect recall ($\ATLKP$)}

The syntax of $\ATLKP$ formulas can be given by the BNF
\[\varphi,\psi::=\bot\mid p\mid (\varphi\Rightarrow\psi)\mid\dknows_\Gamma\varphi\mid\atlD{\Gamma}\circ\varphi\mid\atlD{\Gamma}\until\varphi\psi\mid \atlB{\Gamma}\until\varphi\psi\]
Here $\Gamma$ ranges over finite sets of agents, and $p$ ranges over propositional variables. In this paper we exclude the past temporal operators as their presence does not affect the working of our translation. 

An {\em interpreted system} is defined with respect to some given finite set $\Sigma=\{1,\ldots,N\}$ of {\em agents}, and a set of {\em propositional variables} ({\em atomic propositions}) $AP$. There is also an {\em environment} $e\not\in\Sigma$; in the sequel we write $\Sigma_e$ for $\Sigma\cup\{e\}$.
\begin{definition}[interpreted systems]
An {\em interpreted system} for $\Sigma$ and $AP$ is a tuple of the form \begin{equation}\label{interpretedsystem}
\langle\langle L_i:i\in\Sigma_e\rangle,I,\langle\Act_i:i\in\Sigma_e\rangle,t,V\rangle
\end{equation}
where:

\begin{tabular}{l}
$L_i$, $i\in\Sigma_e$, are nonempty sets of {\em local states};
$L_\Gamma$ stands for $\prod\limits_{i\in\Gamma}L_i$, $\Gamma\subseteq\Sigma_e$;\\
elements of $L_{\Sigma_e}$ are called {\em global states};\\
$I\subseteq L_{\Sigma_e}$ is a nonempty set of {\em initial global states};\\
$\Act_i$, $i\in\Sigma_e$, are nonempty sets of {\em actions};
$\Act_\Gamma$ stands for $\prod\limits_{i\in\Gamma}\Act_i$;\\
$t:L_{\Sigma_e}\times\Act_{\Sigma_e}\rightarrow L_{\Sigma_e}$ is a {\em transition} function;\\
$V\subseteq L_{\Sigma_e}\times AP$ is a valuation of the atomic propositions.
\end{tabular}

\noindent
For every $i\in\Sigma_e$ and $l',l''\in L_{\Sigma_e}$ such that $l'_i=l''_i$ and $l'_e=l''_e$ the function $t$ satisfies $(t(l',a))_i=(t(l'',a))_i$.
\end{definition}
In the literature an interpreted system also includes a {\em protocol} to specify the actions which are permitted at every particular state. Protocols are not essential to our study here as the effect of a prohibited action can be set to that of some fixed permitted action (which is always supposed to exist) to produce an equivalent system in which all actions are always permitted.
Our variant of interpreted systems is borrowed from \cite{LR06a} and has a technically convenient feature which is not present in other works \cite{HFMV95,MCMAS}: every agent's next local state can be directlty affected by the local state of the environment through the transition function. Here follow the technical notions that are relevant to satisfaction of $\ATL$ formulas on interpreted systems.
\begin{definition}[global runs and local runs] 
Given an $n\leq\omega$, a {\em run of length $n$} is a sequence
\[r=l^0a^0l^10a^1\ldots \in L_{\Sigma_e}(\Act_{\Sigma_e}L_{\Sigma_e})^n\]
such that $l^0\in I$ and $l^{j+1}=t(l^j,a^j)$ for all $j<n$. A run is {\em infinite}, if $n=\omega$; otherwise it is {\em finite}. In either case we write $|r|$ for the {\em length} $n$ of $r$. (Note that a run of length $n<\omega$ is indeed a sequence of $2n+1$ states and actions.)

Given $r$ as above and $\Gamma\subseteq\Sigma$, we write 
$r_\Gamma$ for the corresponding {\em local run}
\[l^0_\Gamma a^0_\Gamma \ldots a^{n-1}_\Gamma l^n_\Gamma\in L_\Gamma(\Act_\Gamma L_\Gamma)^n\]
of $\Gamma$ in which $l^j_\Gamma=\langle l^j_i:i\in\Gamma\rangle$ and $a^j_\Gamma=\langle a^j_i:i\in\Gamma\rangle$.

We denote the set of all runs of some fixed length $n\leq\omega$, the set of all finite runs, and the set of all runs in $IS$ by $R^n(IS)$, $R^\fin(IS)$ and $R(IS)$, respectively.

Given $i,j<\omega$ and an $r$ as above such that $i\leq j\leq |r|$, we write $r[i..j]$ for $l^ia^i\ldots a^{j-1}l^j$.
\end{definition}
\begin{definition}[indiscernibility]
Given $r',r''\in R(IS)$ and $i\leq|r'|,|r''|$, we write $r'\sim_{\Gamma,i} r''$ if $r'[0..i]_\Gamma=r''[0..i]_\Gamma$. We write $r'\sim_\Gamma r''$ for the conjunction of $r'\sim_{\Gamma,|r'|} r''$ and $|r'|=|r''|$.
\end{definition}
Sequences of the form $r_\emptyset$ consist of $\langle\rangle$s, and,
consequently, $[r]_\emptyset$ is the class of all the runs of length $|r|$.
Obviously $\sim_{\Gamma,n}$ and $\sim_{\Gamma}$ are equivalence relations on $R(IS)$. 
\begin{definition}
We denote $\{r'\in R(IS): r'\sim_\Gamma r\}$ by $[r]_\Gamma$. 
\end{definition}
\begin{definition}[coalition strategies]
A {\em strategy} for $\Gamma\subseteq\Sigma$ is a vector $s=\langle s_i:i\in\Gamma\rangle$ of functions $s_i$ of type $\{r_i:r\in R^\fin(IS)\}\rightarrow\Act_i$.
We write $S(\Gamma,IS)$ for the set of all the strategies for $\Gamma$ in the considered interpreted system $IS$. Given $s\in S(\Gamma,IS)$ and $r\in R^\fin(IS)$, we write $\out(r,s)$ for the set 
\[\{r'=l^0a^0\ldots a^{n-1}l^n\ldots\in R^\omega(IS):
r'[0..|r|]=r,a^j_i=s_i(r[0..j]_\Gamma)\mbox{ for all }i\in\Gamma\mbox{ and }j\geq |r|\}.\]
of the {\em outcomes} of $r$ when $\Gamma$ sticks to $s$ from step $|r|$ on.
Given an $X\subseteq R^\fin(IS)$, $\out(X,s)$ is $\bigcup\limits_{r\in X}\out(r,s)$.
\end{definition}
Strategies, as defined above, are determined by the local views of the considered coalition members and are therefore {\em uniform}.
\begin{definition}[modelling relation of $\ATLKP$]
The relation $IS,r\models\varphi$ is defined for $r\in R^\fin(IS)$ and formulas $\varphi$ by the clauses:

\noindent
\begin{tabular}{lll}
$IS,r\not\models\bot$;\\
$IS,l^0a^0\ldots a^{n-1}l^n\models p$ & iff & $V(l^n,p)$ for atomic propositions $p$;\\
$IS,r\models\varphi\Rightarrow\psi$ & iff & either $IS,r\not\models\varphi$ or $IS,r\models\psi$;\\
$IS,r\models \dknows_\Gamma\varphi$ & iff & $IS,r'\models\varphi$ for all $r'\in[r]_\Gamma$;\\
$IS,r\models\atlD{\Gamma}\circ\varphi$ & iff & there exists an $s\in S(\Gamma,IS)$ such that \\
& & $IS,r'[0..|r|+1]\models\varphi$ for all $r'\in\out([r]_\Gamma,s)$;\\
$IS,r\models\atlD{\Gamma}\until\varphi\psi$ & iff & there exists an $s\in S(\Gamma,IS)$ s. t. for every $r'\in\out([r]_\Gamma,s)$ there exists \\
& & a $k<\omega$ s. t. $IS,r'[0..|r|+i]\models\varphi$ for all $i<k$ and $IS,r'[0..|r|+k]\models\psi$;\\
$IS,r\models\atlB{\Gamma}\until\varphi\psi$ & iff & for every $s\in S(\Gamma,IS)$ there exist an $r'\in\out([r]_\Gamma,s)$ and a $k<\omega$ s. t.\\
& & $IS,r'[0..|r|+i]\models\varphi$ for all $i<k$ and $IS,r'[0..|r|+k]\models\psi$.
\end{tabular}

\noindent
Validity of formulas in entire interpreted systems and on the class of all interpreted systems, that is, in the logic $\ATLKP$, is defined as satisfaction at all $0$-length runs in the considered interpreted system, and at all the $0$-length runs in all the systems in the considered class, respectively.
\end{definition}
In this paper we assume that each coalition member uses only its own observation power in following a coalition strategy. Allowing coalition members to share their observations gives rise to a more general form of strategy, which are functions of type $\{r_\Gamma:r\in R^\fin(IS)\}\rightarrow\Act_\Gamma$, and which was assumed by the model-checkig algorithm proposed in \cite{GuelevDE11}. 

\subsubsection*{\em Abbreviations}
$\top$, $\neg$, $\vee$, $\wedge$ and $\Leftrightarrow$ have their usual meanings. To keep the use of $($ and $)$ down, we assume that unary connectives bind the strongest, the binary modalities $\atlD{\Gamma}\until..$ and $\atlB{\Gamma}\until..$, and the derived ones below, bind the weakest, and their parentheses are never omitted, and the binary boolean connectives come in the middle, in decreasing order of their binding power as follows: $\wedge$, $\vee$, $\Rightarrow$ and $\Leftrightarrow$. We enumerate coalitions without the $\{$ and $\}$. E.g., the shortest way to write $\atlD{\{1\}}\until{((p\Rightarrow q)\wedge \poss_{\{1\}} r)}{\dknows_{\{2,3\}}(r\vee q))}$ is $\atlD{1}\until{(p\Rightarrow q)\wedge \poss_1 r}{\dknows_{2,3}(r\vee q)}$.
We write $\poss$ for the dual of $\dknows$:
\[\poss_\Gamma\varphi\rightleftharpoons\neg\dknows_\Gamma\neg\varphi.\]
The rest of the combinations of the cooperation modality and future temporal connectives are defined by the clauses
\[\begin{array}{lll}
\atlD{\Gamma}\Diamond\varphi\rightleftharpoons\atlD{\Gamma}\until\top\varphi 
&
\atlD{\Gamma}\Box\varphi\rightleftharpoons\neg\atlB{\Gamma}\Diamond\neg\varphi
&
\atlD{\Gamma}\wuntil\varphi\psi\rightleftharpoons\neg\atlB{\Gamma}\until{\neg\psi}{\neg\psi\wedge\neg\varphi}
\\
\atlB{\Gamma}\Diamond\varphi\rightleftharpoons\atlB{\Gamma}\until\top\varphi
&
\atlB{\Gamma}\Box\varphi\rightleftharpoons\neg\atlD{\Gamma}\Diamond\neg\varphi
&
\atlB{\Gamma}\wuntil\varphi\psi\rightleftharpoons\neg\atlD{\Gamma}\until{\neg\psi}{\neg\psi\wedge\neg\varphi}
\end{array}\]

\subsection{$\ATLKP$ with epistemic objectives only}

In \cite{GuelevDE11} we axiomatized a subset of $\ATLKP$ with past in which $\atlD{.}\until..$ was allowed only in the derived construct $\atlD{\Gamma}\Diamond\dknows_\Gamma\varphi$, and $\atlB{.}\until..$ was allowed only in the derived construct $\atlD{\Gamma}\Box\varphi$. Because of the validity of the equivalences
\[\atlD{\Gamma}\circ\varphi\Leftrightarrow\atlD{\Gamma}\circ\dknows_\Gamma\varphi\mbox{ and }\atlD{\Gamma}\Box\varphi\Leftrightarrow\atlD{\Gamma}\Box\dknows_\Gamma\varphi,\]
that entailed that all the objectives allowed in that subset were epistemic. We argued that, under some assumptions, any $\atlD{.}\until..$ formula could be transformed into an equivalent one of the form $\atlD{\Gamma}\Diamond\dknows_\Gamma\varphi$ thus asserting the significance of the considered subset. Both the axiomatization and the reduction to epistemic goals relied on the presence of the past operators. In this paper we consider another subset of $\ATLKP$. Its formulas have the syntax
\begin{equation}\label{atlsubset}
\varphi,\psi::=\bot\mid p\mid(\varphi\Rightarrow\psi)\mid\dknows_\Gamma\varphi\mid\atlD{\Gamma}\circ\varphi\mid\atlD{\Gamma}\until{\dknows_\Gamma\varphi}{\dknows_\Gamma\psi}
\end{equation}
Unlike the subset from \cite{GuelevDE11}, here we allow formulas of the form $\atlD{\Gamma}\until{\dknows_\Gamma\varphi}{\dknows_\Gamma\psi}$. However, we exclude even the special case $\atlD{\Gamma}\Box\varphi$ of the use of $\atlB{\Gamma}\until{\poss_\Gamma\varphi}{\poss_\Gamma\psi}$. The reasons are discussed in the end of Section \ref{main}.

\subsection{$\CTL$ with distributed knowledge}

This is the target logic of our translation. Its formulas have the syntax
\[
\varphi,\psi::=\bot\mid p\mid(\varphi\Rightarrow\psi)\mid\dknows_\Gamma\varphi\mid\exists\circ\varphi\mid\exists\until\varphi\psi\mid\forall\until\varphi\psi
\]
where $\Gamma$ ranges over finite sets of agents as above. The clauses for the semantics of the connectives in common with $\ATLKP$ are as in $\ATLKP$; the clauses about formulas built using $\exists$ and $\forall$ are as follows:

\begin{tabular}{lll}
$IS,r\models\exists\circ\varphi$ & iff & there exists an $r'\in R^{|r|+1}(IS)$ such that $r=r'[0..|r|]$ and $IS,r'\models\varphi$;\\
$IS,r\models\exists\until\varphi\psi$ & iff & there exists an $r'\in R^\omega(IS)$ such that $r=r'[0..|r|]$ and a $k<\omega$\\
& & such that $IS,r'[0..|r|+i]\models\varphi$ for all $i<k$ and $IS,r'[0..|r|+k]\models\psi$;\\
$IS,r\models\forall\until\varphi\psi$ & iff & for every $r'\in R^\omega(IS)$ such that $r=r'[0..|r|]$ there exists a $k<\omega$ such that \\
& & $IS,r'[0..|r|+i]\models\varphi$ for all $i<k$ and $IS,r'[0..|r|+k]\models\psi$.
\end{tabular}

\noindent
Note that the the occurrences of $\dknows_\emptyset$ is vital for the validity of the equivalences \[\poss_\emptyset\exists\circ\varphi\Leftrightarrow\atlB{\emptyset}\circ\varphi,\ \ \poss_\emptyset\exists\until\varphi\psi\Leftrightarrow\atlB{\emptyset}\until\varphi\psi\mbox{ and }\dknows_\emptyset\forall\until\varphi\psi\Leftrightarrow\atlD{\emptyset}\until\varphi\psi.\]
in the combined language of $\ATLKP$ and $\CTL$ because of the requirement on strategies to be uniform; e.g., $\atlD{\emptyset}\until\varphi\psi$ means that $\until\varphi\psi$ holds along all the extensions of all the runs {\em which are indiscernible from the reference run to the empty coalition.} Therefore here $\atlD{\emptyset}$ does not subsume $\forall$ in the straightforward way known about the case $\ATL$ of complete information.

The combination $\forall\circ$ and the combinations of $\exists$ and $\forall$ with the derived temporal connectives $\wuntil..$, $\Diamond$ and $\Box$ are defined in the usual way.

\section{A validity preserving translation into $\CTL+\dknows$ with perfect recall}\label{main}

\oomit{
Next we present the way we eliminate the $\ATL$-specific constructs from formulas in the subset (\ref{atlsubset}) of $\ATLKP$. We first do formulas of the form $\atlD{\Gamma}\until{\dknows_\Gamma\varphi}{\dknows_\Gamma\psi}$. Their elimination causes some $\atlD{.}\circ$-subformulas to be introduced. We explain how to eliminate $\atlD{.}\circ$-subformulas once all the occurrences of $\atlD{.}\until..$-objectives have been dealt with, thus obtaining a purely $\CTL+\dknows$ formula which is satisfiable iff the originally given $\ATLKP$ one is. Our result applies to $\ATLKP$ on interpreted systems with {\em finite branching}. This means finitely many initial states and finitely many successors to every global state. We conclude this section by proving that our translation can handle the entire language of $\ATL$ in the case of complete information and discuss the reason for leaving $\atlB{\Gamma}\until..$-subformulas out in the case of incomplete information.

}

Our translation captures the subset of $\ATL$ which is given by the BNF
\[\varphi,\psi::=\bot\mid p\mid(\varphi\Rightarrow\psi)\mid\yesterday\varphi\mid\since\varphi\psi\mid\dknows_\Gamma\varphi\mid\atlD{\Gamma}\circ\varphi\mid\atlD{\Gamma}\until{\dknows_\Gamma\varphi}{\dknows_\Gamma\psi}\]
We explain how to eliminate occurrences of $\atlD{.}$ in formulas of the form $\atlD{\Gamma}\until{\dknows_\Gamma\varphi}{\dknows_\Gamma\psi}$ first. In the sequel we write $[\alpha/p]\beta$ for the substitution of the occurrences of atomic proposition $p$ in $\beta$ by $\alpha$.  
\begin{proposition}\label{elimuntil}
Assuming that $p$ and $q$ are fresh atomic propositions, the satisfiability of \\ $[\atlD{\Gamma}\until{\dknows_\Gamma\varphi}{\dknows_\Gamma\psi}/p]\chi$ (at a $0$-length run) is equivalent to the satisfiability of 
\begin{equation}\label{untilelim}
\begin{array}{lll}
\chi&\wedge&\dknows_\emptyset\forall\Box(p\vee q\Rightarrow\dknows_\Gamma\psi\vee(\dknows_\Gamma\varphi\wedge\atlD{\Gamma}\circ q))\\
&\wedge&\dknows_\emptyset\forall\Box(p\Leftrightarrow\dknows_\Gamma\psi\vee(\dknows_\Gamma\varphi\wedge\atlD{\Gamma}\circ p))\\
&\wedge&\dknows_\emptyset\forall\Box(p\Rightarrow \dknows_\Gamma\psi\vee(\dknows_\Gamma\varphi\wedge\forall\circ
\forall\until{q\Rightarrow\dknows_\Gamma\varphi}{q\Rightarrow\dknows_\Gamma\psi})).
\end{array}
\end{equation}
\end{proposition}

\oomit{
Note that, e.g., using just $p$ with the assignment of $p\vee q$ as in the proposition and removing the second conjunctive member from (\ref{untilelim}) would render the elimination step above incorrect because of the possibility to have an infinite ascending sequence of $\dknows_\Gamma\varphi$-runs consisting of transitions in which $\Gamma$ is not targetting $\until{\dknows_\Gamma\varphi}{\dknows_\Gamma\psi}$ and without a $\dknows_\Gamma\psi$-run among them. A similar problem is avoided by the use of the globally optimal strategy $s$ in the proof.
}

Next we explain how to eliminate occurrences of the "basic" $\ATL$ construct $\atlD{\Gamma}\circ\varphi$. Let $IS$ stand for some arbitrary interpreted system (\ref{interpretedsystem}) with finite branching, with $\Sigma=\{1,\ldots,N\}$ as its set of agents, $AP$ as its vocabulary. We adapt the following simple observation, which works in case $\Act_i$, $i\in\Sigma$ are fixed. Readers who are familiar with the original semantics of $\ATL$ on {\em alternating transition systems} ($\mathit{ATS}$) from \cite{AHK97} will recognize the similarity of our technique with the transformation of {\em concurrent game structures} into equivalent $\mathit{ATS}$ from \cite{GJ04}. Assuming that $\Act_i$, $i\in\Sigma_e$, are pairwise disjoint, and disjoint with $AP$, we consider the vocabulary $AP^\Act = AP\cup \bigcup\limits_{i\in\Sigma_e}\Act_i$.
\begin{definition}
Given $IS$ and $*\not\in\bigcup\limits_{i\in\Sigma_e}\Act_i$, we define the interpreted system 
\[IS^\Act=\langle\langle L_i^\Act:i\in\Sigma_e\rangle,I^\Act,\langle\Act_i:i\in\Sigma_e\rangle,t^\Act,V^\Act\rangle\]
by putting:
\[\begin{array}{llllll}
L_i^\Act & = & L_i\times(\Act_i\cup\{*\}),\ i\in\Sigma_e;\\
I^\Act & = & \{\langle\langle l_i,*\rangle:i\in\Sigma_e\rangle:l\in I\};\\ t^\Act(\langle\langle l_i,a_i\rangle:i\in\Sigma_e\rangle,b) & = & \langle\langle (t(l,b))_i,b_i\rangle:i\in\Sigma_e\rangle;\\
V^\Act(\langle\langle l_i,a_i\rangle:i\in\Sigma_e\rangle,p) & \leftrightarrow & V(\langle l_i,:i\in\Sigma_e\rangle,p)\mbox{ for }p\in AP;\\
V^\Act(\langle\langle l_i,a_i\rangle:i\in\Sigma_e\rangle,b) & \leftrightarrow & b=a_j\mbox{ for }b\in\Act_j,\ j\in\Sigma_e.\\
\end{array}\]
\end{definition}
In short, an $IS^\Act$ state is an $IS$ state augmented with a record of the actions which lead to it, the dummy symbol $*$ being used in initial states. 
Let $R\subseteq L^\Act_{\Sigma_e}\times L^\Act_{\Sigma_e}$ and
$R(\langle\langle l_i,a_i\rangle:i\in\Sigma_e\rangle,\langle\langle v_i,b_i\rangle:i\in\Sigma_e\rangle)\mbox{ iff }v=t^\Act(l,b)$.
Then
$IS^\Act,r\models\exists\circ\varphi$ iff $IS^\Act,r\,a\,l'\models\varphi$ for some $l'\in R(l)$ and the only $a\in\Act_{\Sigma_e}$ such that $r\,a\,l'\in R^\fin(IS^\Act)$. 
The key observation in our approach is that
\begin{equation}\label{nexttimetranslation}
IS,r\models\atlD{i_1,\ldots,i_k}\circ\varphi\mbox{ iff }IS^\Act,r^\Act\models\bigvee\limits_{a_{i_1}\in\Act_{i_1}}\ldots\bigvee\limits_{a_{i_k}\in\Act_{i_k}}\dknows_{\{i_1,\ldots,i_k\}}\forall\circ\left(\bigwedge\limits_{j=1}^ka_{i_j}\Rightarrow\varphi\right)
\end{equation}
For this observation to work without refering to the actions in the particular interpreted system, given an arbitrary $IS$, we enrich it with dedicated actions which are linked to the objectives occurring in the considered formula. We define the transition function on these actions so that if a particular $\circ\varphi$-objective can be achieved at finite run $r$ at all, then it can be achieved by taking the corresponding dedicated actions at the last state of $r$. This can be achieved in forest-like systems where runs can be determined from their final states. Similarly, we introduce express actions for the environment that enable it to foil objectives at states at which they objectives cannot be achieved by the respective coalitions using any strategy based on the original actions. (Giving the environment such powers does not affect the satisfaction of formulas as it never participates in coalitions.) The sets $\Act_i$, $i\in\Sigma_e$ of atomic propositions by which we model actions satisfy the formula
\[\mathsf{A}(\Act_1,\ldots,\Act_N,\Act_e)\rightleftharpoons\bigwedge\limits_{a_1\in\Act_1}\ldots\bigwedge\limits_{a_N\in\Act_N}\bigwedge\limits_{a_e\in\Act_e}\exists\circ\bigwedge\limits_{i\in\Sigma_e}a_i,\]
which states that any vector of actions from $\Act_{\Sigma_e}$ produces a transition. Consider an $\ATLKP$ formula of the form below with no occurrences of $\until..$-objectives:
\begin{equation}\label{elimform}
\chi\wedge\dknows_\emptyset\forall\Box\mathsf{A}(\Act_1,\ldots,\Act_N,\Act_e)
\end{equation}
Here $\Act_1,\ldots,\Act_N,\Act_e$ consist of the atomic propositions which have been introduced to eliminate $\atlD{\Gamma}\circ\varphi$-subformulas so far. For the original $\chi$ we assume $\Act_i=\{\mathsf{nop}_i\}$, $i\in\Sigma_e$, where $\mathsf{nop}_i$ have no specified effect. We remove the occurrences of $\atlD{\Gamma}\circ\varphi$-subformulas in $\chi$ working bottom-up as follows.
\begin{proposition}\label{elimcircprop}
Let $\mathsf{a}_{\Gamma,i,\varphi}$, $i\in\Gamma\cup\{e\}$, be fresh atomic propositions,
$\Act_i'=\Act_i\cup\{\mathsf{a}_{\Gamma,i,\varphi}\}$ for $i\in\Gamma\cup\{e\}$ and $\Act_i'=\Act_i$ for $i\in\Sigma\setminus\Gamma$.
Then the satisfiability of  
\begin{equation}\label{elimcirc0}
[\atlD{\Gamma}\circ\varphi/p]\chi\wedge\dknows_\emptyset\forall\Box\mathsf{A}(\Act_1,\ldots,\Act_N,\Act_e)
\end{equation}
entails the satisfiability of the formula
\begin{equation}\label{elimcirc}
\begin{array}{l}
\left[\dknows_\Gamma\forall\circ\left(\bigwedge\limits_{i\in\Gamma}\mathsf{a}_{\Gamma,i,\varphi}\Rightarrow\varphi\right)/p\right]\chi\wedge\\
\dknows_\emptyset\forall\Box\left(\dknows_\Gamma\forall\circ\left(\bigwedge\limits_{i\in\Gamma}\mathsf{a}_{\Gamma,i,\varphi}\Rightarrow\varphi\right)\vee
\poss_\Gamma\forall\circ(\mathsf{a}_{\Gamma,e,\varphi}\Rightarrow\neg\varphi)\right)
\wedge\\
\dknows_\emptyset\forall\Box\mathsf{A}(\Act_1',\ldots,\Act_N',\Act_e').
\end{array}
\end{equation}
\end{proposition}
\oomit{To realize that the satisfiability of a $\CTL+\dknows$ formula of the form (\ref{elimform}) obtained as described above entails the satisfiability of the original $\ATLKP$ one, consider a satisfying structure of the form  \begin{equation}\label{ctldis}
IS^-=\langle\langle L_i:i\in\Sigma_e\rangle,I,-,V\rangle
\end{equation}
where $-$ represents the passage of time and choose $t(l,a)\in -(l)\cap\bigcap\limits_{i\in\Sigma_e}\{l'\in L_{\Sigma_e}:V(l',a_i)\}$,  $a\in\Act_{\Sigma_e}$, $l\in L_{\Sigma_e}$. The nonemptyness of the latter set is guaranteed by the validity of $\mathsf{A}(\Act_1,\ldots,\Act_N,\Act_e)$.}
The above proposition shows how to eliminate one by one all the occurrences of the cooperation modalities in an any given $\ATLKP$ formula $\chi$ with the cooperation modalities appearing only in subformulas of the form $\atlD{\Gamma}\circ\varphi$ and obtain a $\CTL+\dknows$ formula $\chi'$ such that if $\chi$ is satisfiable, then so is $\chi'$. Now consider a purely-$\CTL+\dknows$ formula of the form (\ref{elimform}). The satisfaction of (\ref{elimform}) requires just a transition relation for the passage of time to define as it contains no $\atlD{\Gamma}$s and hence no reference to actions. That is, we assume a satisfying model of the form
\begin{equation}\label{ctldis}
IS^-=\langle\langle L_i:i\in\Sigma_e\rangle,I,-,V\rangle
\end{equation}
where $L_i$, $i\in\Sigma_e$, $I$ and $V$ are as in interpreted systems, and $-$ is a serial binary relation on the set of the global states $L_{\Sigma_e}$ that represents the passage of time. We define the remaining interpreted system components as follows. We choose the set of actions of each agent $i$, including the environment, to be the corresponding set of atomic propositions $\Act_i$ from (\ref{elimform}). For any $a\in\Act_{\Sigma_e}$ and any $l\in L_{\Sigma_e}$ we choose $t(l,a)$ to be an arbitrary member of $-(l)\cap\bigcap\limits_{i\in\Sigma_e}\{l'\in L_{\Sigma_e}:V(l',a_i)\}$. The nonemptiness of the latter set is guaranteed by the validity of $\mathsf{A}(\Act_1,\ldots,\Act_N,\Act_e)$ in $IS^-$, which states that every state has a successor satisfying the conjunction $\bigwedge\limits_{i\in\Sigma_e} a_i$ for any given vector of actions $a\in\Act_{\Sigma_e}$. Let $IS$ stand for the system obtained by this definition of $\Act_i$, $i\in\Sigma_e$, and $t$. It remains to show that \begin{equation}\label{elimequiv}
IS,r\models\dknows_\Gamma\forall\circ\left(\bigwedge\limits_{i\in\Gamma}\mathsf{a}_{\Gamma,i,\varphi}\Rightarrow\varphi\right)
\end{equation}
is equivalent to $IS,r\models\atlD{\Gamma}\circ\varphi$ for any subformula $\atlD{\Gamma}\circ\varphi$ eliminated in the process of obtaining (\ref{elimform}). For the forward direction, establishing that the actions ${a}_{\Gamma,i,\varphi}$, $i\in\Gamma$ provides $\Gamma$ with a strategy to achieve $\varphi$ in one step is easily done by a direct check. For the converse direction, if (\ref{elimequiv}) is false, then the validity of the second conjunctive member of (\ref{elimcirc}) entails that $\Gamma$ cannot rule out the possibility that the environment can enforce $\neg\varphi$ in one step by choosing its corresponding action $\mathsf{a}_{\Gamma,e,\varphi}$.

\subsubsection*{Formulas of the form $\atlB{\Gamma}\until{\poss_\Gamma\varphi}{\poss_\Gamma\psi}$}
We first note that no restriction on formulas of the respective more general form $\atlB{\Gamma}\until\varphi\psi$ is necessary in the case of complete information. 
\begin{proposition}[eliminating $\atlB{\Gamma}\until\varphi\psi$ in $\ATL$ with complete information]\label{atlbcompleteinformation}
Let $p$ and $q$ be some fresh atomic propositions. The satisfiability of
\[[\atlB{\Gamma}\until\varphi\psi/p]\chi\]
in $\ATL$ with complete information is equivalent to the satisfiability of
\begin{equation}\label{buntilelim}
\begin{array}{lll}
\chi&\wedge&\forall\Box(p\vee q\Rightarrow\psi\vee(\varphi\wedge\atlB{\Gamma}\circ q))\\
&\wedge&\forall\Box(p\Leftrightarrow\psi\vee(\varphi\wedge\atlB{\Gamma}\circ p))\\
&\wedge&\forall\Box(p\Rightarrow\psi\vee(\varphi\wedge\forall\circ
\forall\until{q\Rightarrow\varphi}{q\Rightarrow\psi})).
\end{array}
\end{equation}
\end{proposition}
In the incomplete information case our approach suggests replacing  $[\atlB{\Gamma}\until{\poss_\Gamma\varphi}{\poss_\Gamma\psi}/p]\chi$ by
\[\begin{array}{lll}
\chi&\wedge&\dknows_\emptyset\forall\Box(p\vee q\Rightarrow\poss_\Gamma\psi\vee(\poss_\Gamma\varphi\wedge\atlB{\Gamma}\circ q))\\
&\wedge&\dknows_\emptyset\forall\Box(p\Leftrightarrow\poss_\Gamma\psi\vee(\poss_\Gamma\varphi\wedge\atlB{\Gamma}\circ p))\\
&\wedge&\dknows_\emptyset\forall\Box(p\Rightarrow \poss_\Gamma\psi\vee(\poss_\Gamma\varphi\wedge\ldots)).
\end{array}\]
where, in a forest-like system $IS$, $q$ is supposed to mark states which are reached from runs $r$  in which $\Gamma$ cannot achieve $\until{\poss_\Gamma\varphi}{\poss_\Gamma\psi}$ when $\Gamma$'s actions $a$ are complemented on behalf of the non-members of $\Gamma$ by some actions $b_{a_1,r_1}$ that foil the objective, and $\ldots$ is supposed to express that any sequence of vectors of actions $a_1,a_2,\ldots\in\Act_\Gamma$ when complemented by the corresponding $b_{a_1,r_1}$, $b_{a_2,r_2},\ldots$ can generate a sequence $r_1,r_2,\ldots$ of finite runs, starting with the reference one, each of them being $\Gamma$-indiscernible from the extension of the previous one, by the outcome of the respective $a_k\cdot b_{a_k,r_k}$, such that there exists a $k<\omega$ with $IS,r_j\models q\wedge\dknows_\Gamma\varphi$, $j=1,\ldots,k-1$, and $IS,r_k\models \neg q\vee\dknows_\Gamma\psi$. The  fixpoint construct that would best serve expressing this condition can be written as $\mu X.\alpha\vee(\beta\wedge\poss_\Gamma\forall\circ X)$ in the modal $\mu$-calculus (cf. e.g. \cite{HandbookModalLogicBradfield}). Finding a substitute for it in $\CTL+\dknows$ appears problematic. 
 
\section*{Concluding remarks}

\oomit{The way $\CTL$ path quantifiers are generalized in $\ATL$ is well known in the literature. In this paper we have shown how this connection facilitates automating proof in $\ATL$. We have established the a reduction of $\ATL$ validity to $\CTL$ validity in the case of an epistemic extension showing that a corresponding epistemic extension of $\CTL$ can be used as the target logic. The key step in our reduction is the elimination of $\atlD{.}\circ$-subformulas in a manner that is independent from the possible sets of actions. We achieve this by adapting an element of the the completeness arguments for $\ATL$ axiomatic systems \cite{GuelevDE11,GD06}: we show that any model can be augmented by dedicated actions which can be used as a winning strategy for their corresponding $\circ$-objective provided that the model admits a winning strategy for the objective at all. Similar {\em witness} actions play a central role in satisfying model constructions. Since these auxiliary dedicated actions bear the same names in all models, making explicit mention of the atomic propositions which we use to model them in $\CTL$ does not bind the meaning of the obtained $\CTL$ formula to a particular model.
}

Our approach is inspired by temporal resolution \cite{DBLP:journals/tocl/FisherDP01}, which has been extended to epistemic $\LTL$ \cite{DBLP:journals/logcom/DixonFW98} and to (non-epistemic) $\CTL$ and $\CTL^*$ \cite{DBLP:journals/jetai/BolotovF99,DBLP:conf/mfcs/BolotovDF99}, the latter system being the closest to our target system $\CTL+\dknows$. Following the example of these works, a resolution system for $\CTL+\dknows$ could be proved complete by showing how to reproduce in it any proof in some complete, e.g., Hilbert style proof system. A complete axiomatization for epistemic $\CTL^*$ with perfect recall can be found in \cite{MeydenWong03}, but the completeness was demonstrated with respect to the so-called {\em bundle} semantics, where a model may consist of some set of runs that need not be all the runs generated by a transition system. and the form of collective knowledge considered in \cite{MeydenWong03} is {\em common knowledge}, whereas we have distributed knowledge. The setting for the complexity results from \cite{DBLP:conf/stoc/HalpernV86} is similar. The tableau-based decision procedure for epistemic $\CTL$ with both common and distributed knowledge from \cite{DBLP:conf/mallow/GorankoS09} does not cover the case of perfect recall. To the best of our knowledge no decision procedure of feasible complexity such as the resolution- and tableau-based ones that are available for so many closely related systems from the above works has been developed yet for validity in $\CTL+\dknows$ with perfect recall.

\section*{Acknowledgement}
The research in this paper was partially supported through Bulgarian National Science Fund Grant DID02/32/2009.

\bibliographystyle{alpha}
\bibliography{../../bibfiles/mybiblio}

\end{document}